\RequirePackage{fix-cm}
\documentclass[smallextended]{svjour3}       
\smartqed  
\usepackage{graphicx}
\usepackage{blindtext}
\usepackage{amsmath}
\usepackage{mathptmx}      
\usepackage{cite}
\usepackage{lineno}
\usepackage[numbers]{natbib}
\usepackage{amsmath}
\usepackage{alphabeta} 
\usepackage{appendix}
\usepackage{chngcntr}

\journalname{Journal of Materials Research}
\begin{document}

\title{Calibration and data analysis routines for nanoindentation with spherical tips}


\author{Diana Avadanii         \and
        Anna Kareer    \and
        Lars Hansen  \and
        Angus Wilkinson
}

\institute{D. Avadanii \at
              Department of Earth Sciences, University of Oxford \\
              \email{diana.avadanii@univ.ox.ac.uk} \\
           \and
           A. Kareer \at
              Department of Materials,
              University of Oxford \\
              \email{ anna.kareer@materials.ox.ac.uk} \\
           \and
            L. Hansen \at
              Department of Earth and Environmental Sciences,
              University of Minnesota \\
            \and
            A. Wilkinson \at
              Department of Materials,
              University of Oxford
}

\date{Received: date / Accepted: date}

\maketitle
\begin{abstract}

Instrumented spherical nanoindentation with a continuous stiffness measurement has gained increased popularity in material science studies in brittle and ductile materials alike. These investigations span hypotheses related to a wide range of microphysics involving grain boundaries, twins, dislocation densities, ion-induced damage and more. These studies rely on the implementation of different methodologies for instrument calibration and for circumventing tip shape imperfections. In this study, we test, integrate, and re-adapt published strategies for tip and machine-stiffness calibration for spherical tips. We propose a routine for independently calibrating the effective tip radius and the machine stiffness using three reference materials (fused silica, sapphire, glassy carbon), which requires the parametrization of the effective radius as a function of load. We validate our proposed workflow against key benchmarks, such as variation of Young's modulus with depth. We apply the resulting calibrations to data collected in materials with varying ductility (olivine, titanium, and tungsten) to extract indentation stress-strain curves. We also test the impact of the machine stiffness on recently proposed methods for identification of yield stress, and compare the influence of different conventions on assessing the indentation size effect. Finally, we synthesize these analysis routines in a single workflow for use in future studies aiming to extract and process data from spherical nanoindentation.


\end{abstract}

\section{Introduction}
\label{intro-sph-nanoindentation}
Instrumented nanoindentation has become a widely-used technique for material characterization. The array of available tip geometries allows material analysis over a wide range of stress states. Indentation with a rigid, spherical tip has recently gained popularity due to key advantages of this geometry compared to sharp tip geometries (e.g., Berkovich). Indentation with a sphere induces a stress field under the indenter that is not self similar, and therefore the indentation strain progressively increases with indentation depth \citep[see Chapter 3 in][]{fischer2011analysis, pharr2010indentation, pathak2015sphericalreview, johnson1970correlation}. Thus, load-displacement data can be transformed into indentation stress-strain curves, which facilitate the investigation of the full elasto-plastic response of materials \citep[for review, see][]{pathak2015sphericalreview}. Moreover, nanoindentation with indenters of varying radii provides insight into size effects associated with deformation and enables correlation of small-scale material testing with macroscopic tests \citep[e.g.,][]{pathak2015sphericalreview}. For example, in certain materials, indentation using tips with large radii results in stress-strain curves with a hardening coefficient similar to that of uniaxial tests, while data obtained using indenters with smaller radii display an indentation size effect \citep[e.g.,][]{spary2006indentation, weaver2016mechanical}. In spherical nanoindentation, the indentation size effect is underpinned by both material hardening with increasing indentation strain, as well an increase in hardness with decreasing spherical radii \citep{swadener2002sizeeffects, pharr2010indentation, hou2008study, herbert2001measurement, spary2006indentation}. Finally, bursts of displacement in the stress-strain curves (called ‘pop-ins’) provide valuable insight into initiation of plasticity at small scales and display an additional size effect in spherical nanoindentation, in which the stress at pop-in increases with decreasing tip radius \citep{zhu2008size, morris2011popin, shim2008different, patel2020spherical}.

The potential to produce a large number of measurements and generate indentation stress-strain curves from a small volume of material makes spherical nanoindentation a desirable technique with a wide range of applications. For example, spherical nanoindentation has been deployed to investigate yield stress and size effects in brittle engineering ceramics and natural minerals \citep[e.g.,][]{hackett2019evaluation, zhu2008size, kumamoto2017size, hansen2020insight, stich2022room, fang2021nanoindentation}, the mechanical properties of twin and grain boundaries in metals \citep[e.g.,][]{vachhani2016GBstudies, weaver2018quantifying, li2021popin}, the effects of ion-induced damage in metals and alloys \citep[e.g.,][]{bushby2012Feiondamage,armstrong2015ion,pathak2017iondamage}, the relationship between structure and mechanical properties in biomaterials (bone \citep[e.g.,][]{pathak2012bone}, human enamel \citep[e.g.,][]{he2006humanenamel}), elasto-plastic transitions in bulk metallic glasses \citep[e.g,][]{choi2012estimation, bei2004theoretical}, and fracture in thin films \citep[e.g.,][]{zhang2017thinfilm, mercier2017thinfilm}. This versatility of mechanical testing using spherical indentation has motivated studies on the reliability of measurements and prompted efforts to improve data analysis. Investigations into the methodology of spherical indentation highlight the importance of accurate knowledge of tip radius \citep{leitner2018essential, li2013effects, bushby2000areafunction}, machine stiffness, \citep{li2013effects, kang2012, cabibbo2012international}, differences among nanoindenter apparatus \citep{cabibbo2012international}, and the impact of different data-analysis routines on the extracted material properties \citep{pathak2015sphericalreview,kalidindi2008ZPC}. 

In this contribution, we test, integrate, and adapt several published strategies for calibration of spherical tip shape and machine stiffness, and implement a routine to calibrate the effective radius and machine stiffness using purely elastic indentation data collected on materials with varying elastic moduli. This synthesis results in an improved workflow to accurately extract stress-strain curves. We also develop a procedure for characterization of the effective tip shape and machine stiffness for tips that are not perfectly spherical, relying on the obtained indentation data from reference materials. We emphasize that the cumulative impact of imperfections in spherical tips, inconsistencies in sample mounting, and variations among instruments reduce the repeatability and overall meaningful interpretation of data collected across multiple studies. To this end, we outline a calibration routine using fused silica, sapphire, and glassy carbon as reference materials. We then implement the resulting calibration to extract indentation stress-strain curves from tungsten, olivine, and titanium. Details of the experiment set-up are outlined in Section \ref{expMethod}. We suggest that future studies using spherical nanoindentation, would benefit from detailed reports of how the tip and machine-stiffness calibration satisfy different benchmarks for meaningful comparison among published results. Ultimately, we intend for this contribution to serve as a detailed guide to deliver the full potential of spherical nanoindentation as a materials characterisation technique with application to a broad range of materials with varying ductility. 

\section{Results and Analysis}
\label{results&discussion-sphericalnanoind}
\subsection{Theoretical background}
\label{background-sphericalnanoind}

Instrumented nanoindentation measures the displacement of the indenter tip in response to an applied load, $P$. The total measured displacement, $h_\mathrm{total}$, is a combination of the displacement due to surface deformation of the sample, $h_\mathrm{sample}$, and the deflection due to the machine stiffness, $h_\mathrm{mach}$. This interaction can be expressed as two deforming elements connected in series, for which the displacements are \citep{li2002CSMreview, li2002CSMreview, oliverandpharr92, swadener1999calibration}
\begin{equation}
h_\mathrm{total} = h_\mathrm{sample} + h_\mathrm{mach}.
\label{eqn:def_h_total}
\end{equation}
The value of $h_\mathrm{mach}$ can be estimated from the machine stiffness, $S_\mathrm{mach}$, according to  $h_\mathrm{mach} = P/S_\mathrm{mach}$, where the machine stiffness accounts for the combined stiffness from the indenter tip and the loading frame. The common procedure is to determine $S_\mathrm{mach}$ for an instrument by performing a series of experiments with a Berkovich tip in a material with known elastic modulus. This method allows both the area function of the tip and the machine stiffness to be determined simultaneously \citep[e.g.,][]{oliverandpharr92}. The obtained value of the machine stiffness is used as the default stiffness, $S_\mathrm{default}$, which is generally applied by the instrument software when collecting and reporting new data. However, previous studies using spherical indenters have indicated that the stiffness during experiments is a function of the applied load, rather than a single value \citep{li2002CSMreview, cabibbo2012international}. Consequently, this procedure for calibration of the area function and machine stiffness can lead to systematic errors in subsequent experiments if the applied loads are significantly different than those used in the calibration \citep{li2013effects}.

It is possible to explicitly implement a calibration routine for spherical indentation tips in order to identify $S_\mathrm{mach}$ for each machine-tip pair, as well as to determine the effective radius of the tip. Following \citet{li2002CSMreview}, we express the reported displacement as 
\begin{equation}
h_\mathrm{rep} = h_\mathrm{total} - \frac{P}{S_\mathrm{default}}+ h_\mathrm{0}, 
\label{eqn:def_h_rep}
\end{equation}
where $h_\mathrm{0}$ accounts for errors arising from the initial contact between the sample surface and the indenter tip (Figure \ref{fig:Supplementary-cartoon}). 

Because the stiffness of a particular indenter tip combined with the stiffness of the other components of the instrument are unknown,  we can describe $h_\mathrm{rep}$ as
\begin{equation}
h_\mathrm{rep} = h_\mathrm{sample} + \frac{P}{S_\mathrm{mach}} - \frac{P}{S_\mathrm{default}} + h_\mathrm{0}.
\label{eqn:def_h_rep2}
\end{equation}
For cases in which the response of the material is purely elastic, $h_\mathrm{sample}$ can be modelled according to Hertzian mechanics as the elastic displacement, $h_\mathrm{e}$  \citep{johnson1970correlation},
\begin{equation}
    h_\mathrm{e} = P^{2/3}\left ( \frac{4}{3} \sqrt{R_\mathrm{eff}}E_\mathrm{eff} \right )^{-2/3}, 
\label{eqn:h_e}
\end{equation}
for which
\begin{equation}
    \frac{1}{E_\mathrm{eff}} = \frac{1-\upsilon_\mathrm{s}}{E_\mathrm{s}} + \frac{1-\upsilon_\mathrm{i}}{E_\mathrm{i}} 
\label{eqn:E_eff_def}
\end{equation}
and
\begin{equation}
    \frac{1}{R_\mathrm{eff}} = \frac{1}{R_\mathrm{s}} + \frac{1}{R_\mathrm{i}}. 
\label{eqn:R_eff_def}
\end{equation}
$E_\mathrm{eff}$ and $R_\mathrm{eff}$ are the reduced elastic modulus and tip radius, and are expressed as a function of the elastic moduli and Poisson's ratios of the sample ($E_\mathrm{s}$ and $\upsilon_\mathrm{s}$) and indenter tip ($E_\mathrm{i}$ and $\upsilon_\mathrm{i}$) and of the radius of the sample surface ($R_\mathrm{s}$) and indenter tip ($R_\mathrm{i}$). For a purely elastic contact and a flat sample surface, the curvature of the surface is infinity so that $R_\mathrm{eff} = R_\mathrm{i}$ (Figure \ref{fig:Supplementary-cartoon}). 

\subsection{Calibrations on materials with known moduli}
\label{referenceMaterials}
\subsubsection{Determination of effective radius}
\label{radius_calculation}
\begin{figure*}
  \includegraphics[width=0.9\textwidth]{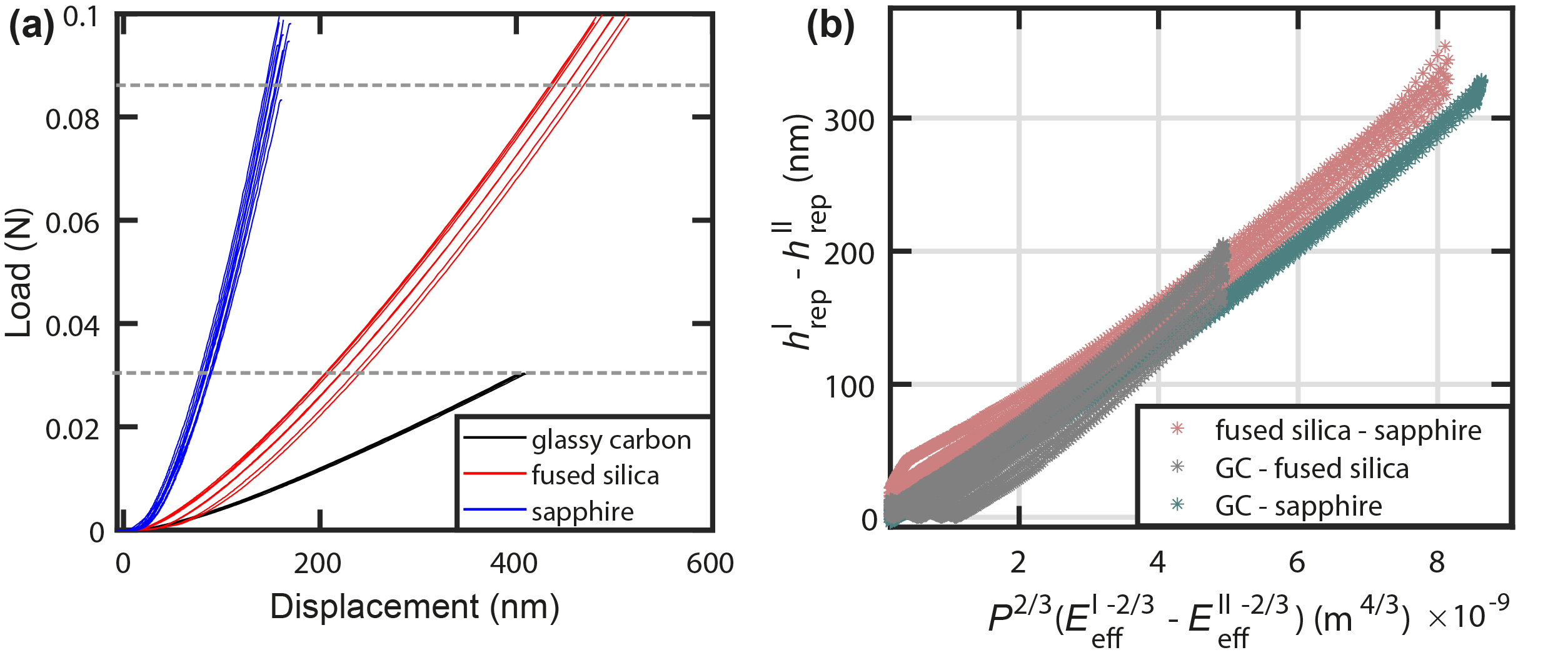}
\caption{a) Load-displacement data collected with a spherical tip with a nominal radius of 10 \textnormal{\mu}m in different materials with known moduli. b) Difference between measured displacements in two materials at similar load. The end cut-off load for each set of experiments is marked by the dashed horizontal lines in a). The slope of these curves is proportional to the effective tip radius according to Equation \ref{eqn:delta_h_rep}.}
\label{figure1}       
\end{figure*}

Studies employing spherical nanoindentation implement various strategies for calculating the effective radius, $R_\mathrm{eff}$ \citep{pathak2015sphericalreview, li2002CSMreview, leitner2018essential,bushby2000areafunction}. The value of $R_\mathrm{eff}$ can be calculated by using Equation \ref{eqn:R_eff_def} and the nominal tip radius provided by the manufacturer \citep{hackett2019evaluation}. More commonly, $R_\mathrm{eff}$ is calculated by fitting Equation \ref{eqn:h_e} to data collected in fused silica \citep{swadener1999calibration, pathak2015sphericalreview, patel2020spherical}. Alternatively, using data from fully elastic experiments in materials with a wide range of elastic moduli, optimum values of $R_\mathrm{eff}$ can be simultaneously determined alongside other variables (e.g., machine stiffness) with the constraint that $R_\mathrm{eff}$ is as constant with depth as possible \citep[e.g.,][]{fieldandswain1993simple, bushby2000areafunction, zhu2008size}. Complications in implementing these methods arise due to the impact of machine stiffness on reported displacement values and of tip shape imperfections, which are difficult effects to deconvolve. These errors can lead to unrealistic differences in the values of $R_\mathrm{eff}$ obtained with different reference materials (e.g., fused silica and sapphire) \citep{li2002CSMreview, bushby2000areafunction, zhu2008size}. Moreover, errors in the calibration of $R_\mathrm{eff}$ can lead to inconsistencies in stress-strain measurements on the same material with different indenter tips. These potential discrepancies motivate the need for a calibration routine in which the machine stiffness and effective radius are both determined in a self-consistent manner \citep{li2002CSMreview}.

To circumvent these issues, we implement the method proposed by \citet{li2013effects} to find the machine stiffness and effective radius for each machine-tip pair in Table \ref{table_experiemntSummary}.
Although \citet{li2013effects} applied their calibration routine to two reference materials, we extend this analysis to three reference materials. This approach relies on the difference in reported displacements at the same load in elastic experiments on different materials with known elastic moduli. Following \citet{li2013effects}, we rearrange Equation \ref{eqn:def_h_rep2} to define the error in displacement, $h_\mathrm{err}$, as
\begin{equation}
h_\mathrm{err} =  h_\mathrm{rep} -  h_\mathrm{e} - h_\mathrm{0} =  \frac{(S_\mathrm{default}-S_\mathrm{mach})P}{S_\mathrm{default}S_\mathrm{mach}}.
\label{eqn:def_h_err}
\end{equation}

The main assumption with this approach is that, at a given load, $h_\mathrm{err}$ is the same for different reference materials. For two reference materials (noted as I and II), using Equation \ref{eqn:def_h_rep2} and data collected at the same load, we can subtract the reported elastic displacement in material II, $h_\mathrm{rep}^\mathrm{II}$, from the reported elastic displacement in material I, $h_\mathrm{rep}^\mathrm{I}$. The $\frac{P}{S_\mathrm{mach}}$ and $\frac{P}{S_\mathrm{default}}$ terms in Equation \ref{eqn:def_h_rep2} corresponding to each material cancel out for the same load $P$ such that \citep{li2013effects}
\begin{equation}
h_\mathrm{rep}^\mathrm{I} - h_\mathrm{rep}^\mathrm{II} = 
P^{2/3}\left ( \frac{4}{3} \sqrt{R_\mathrm{eff}}\right )^{-2/3}\left ({E_\mathrm{eff}^\mathrm{I}}^{-2/3} - {E_\mathrm{eff}^\mathrm{II}}^{-2/3} \right )+ h_\mathrm{0}^\mathrm{I} - h_\mathrm{0}^\mathrm{II}. 
\label{eqn:delta_h_rep}
\end{equation}

We apply Equation \ref{eqn:delta_h_rep} to data collected from fused silica, glassy carbon, and sapphire as outlined in Section \ref{expMethod}. In Figure \ref{figure1}, we display the reported elastic load-displacement curves in the reference materials collected with a tip with nominal radius  $R_\mathrm{n} = 10$ \textnormal{\mu}m, and mark the load and displacement used in Equation \ref{eqn:delta_h_rep} with a dashed horizontal line. Figure \ref{figure1}b reports the differences calculated using Equation \ref{eqn:delta_h_rep} between reported displacement in fused silica and sapphire, glassy carbon and fused silica, and glassy carbon and sapphire as a function of a term proportional to $P^{2/3}$. Thus, according to Equation \ref{eqn:delta_h_rep} the slope of the graph in Figure \ref{figure1}b is proportional to $R_\mathrm{eff}$, and the intersection with the vertical axis is the difference in the displacement error due to surface contact in the two materials ($h_\mathrm{0}^\mathrm{I} - h_\mathrm{0}^\mathrm{II}$). 

\begin{figure*}
  \includegraphics[width=0.5\textwidth]{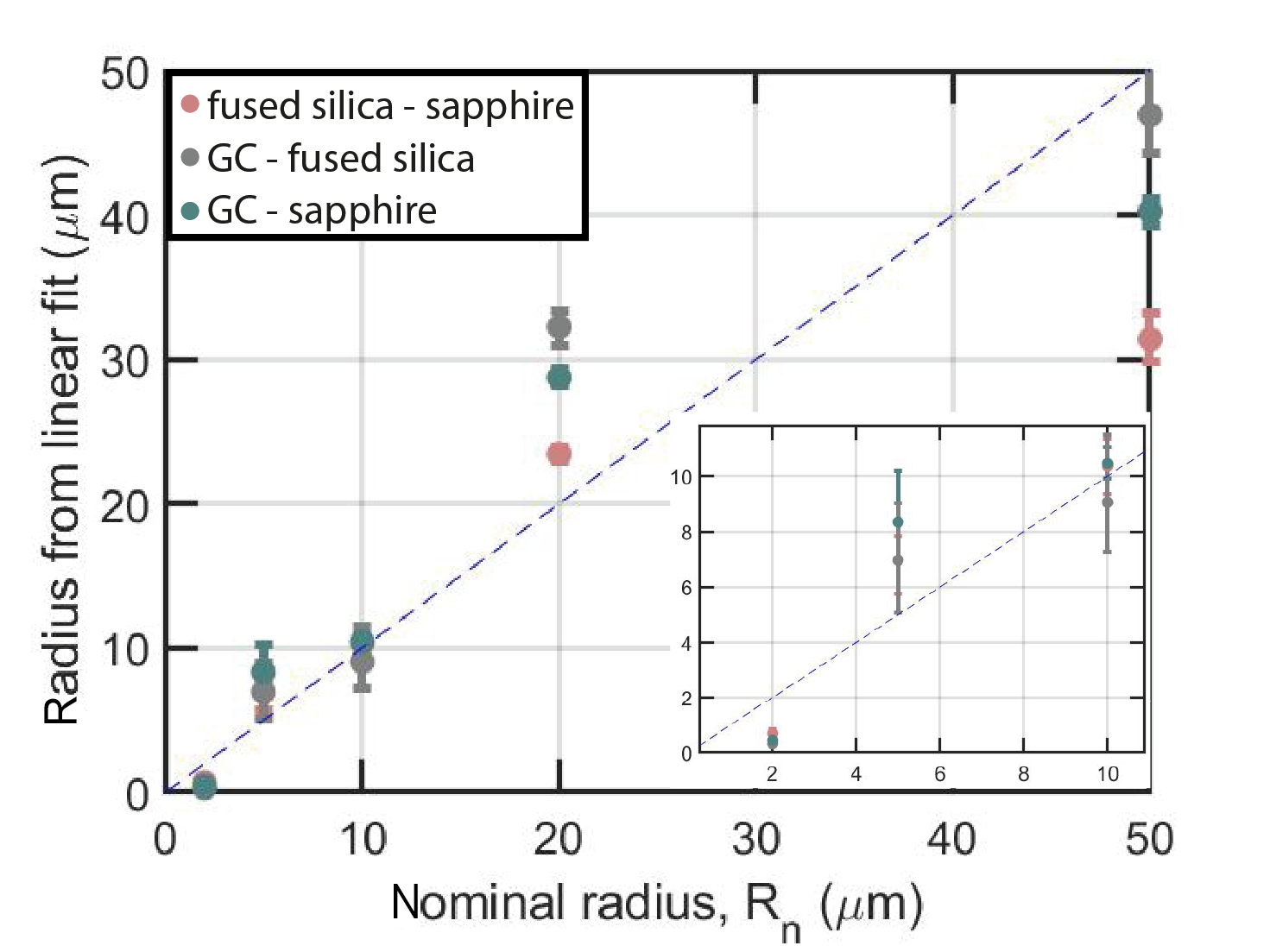}
\caption{Summary of results for effective radius determined as the slope of a line fit through data calculated as the subtraction of elastic load-displacement curves at the same load in different reference materials following the method of \citet{li2013effects} (Equation \ref{eqn:delta_h_rep}). The dashed line represents the 1:1 proportionality.}
\label{figure6}       
\end{figure*}

One key observation in Figure \ref{figure1}b is that the average slope of the curves systematically varies among the three pairs of reference materials. This observation is ubiquitous among our experiments, as can be seen in Figure \ref{figure6}, which demonstrates that the best-fit $R_\mathrm{eff}$ varies by up to $40\%$ depending on the pair of reference materials. This result is contrary to expectations arising from the analysis of \citet{li2013effects}, which suggests that $R_\mathrm{eff}$ should not depend on the reference materials used in calibration.  

\begin{figure*}
  \includegraphics[width=0.9\textwidth]{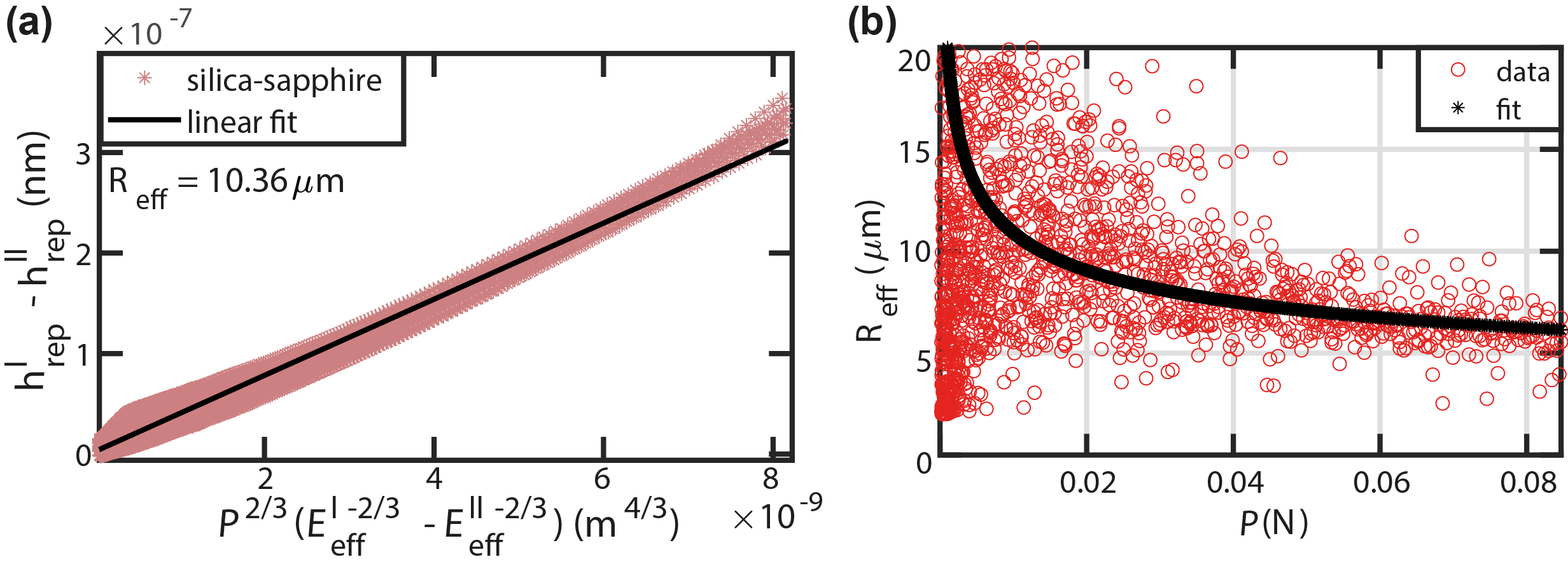}
\caption{a) Linear fit for a spherical tip with a nominal radius of 10 \textnormal{\mu}m following \citet{li2013effects}. Note the curvature in the data obtained by subtracting sapphire displacement from fused silica displacement according to Equation \ref{eqn:delta_h_rep}. b) Effective radius calculated using the first numerical derivative of data in a) against load and an exponential fit. All values for $R_\mathrm{eff}$ in Figure \ref{figure6} and the values for $R_\mathrm{eff}$ at a constant load can be found in Supplementary Materials, Table \ref{table_resultsSummary}.}
\label{figure2}       
\end{figure*}

Part of this discrepancy results from the curves in Figure \ref{figure1}b departing from linearity, which implies that $R_\mathrm{eff}$ is not a constant for any given calibration. This issue is accentuated in Figure \ref{figure2}a, which presents data comparing fused silica to sapphire and a linear fit assuming constant $R_\mathrm{eff}$. There is clearly curvature in the data not captured by the linear fit. Since the subtraction method proposed by \citet{li2013effects} accounts for the effects of machine stiffness when calculating $R_\mathrm{eff}$, we interpret the curvature of the data in Figure \ref{figure2}a to instead result from departure of the tip shape from a perfect sphere. Imperfections in tip shape could be accounted for by the parametrization of $R_\mathrm{eff}$ as a function of displacement. However, measured displacements are also affected by the machine stiffness, which is unknown at this point in the analysis. Therefore, we instead choose to express the effective radius as a function of load. Figure \ref{figure2}b presents $R_\mathrm{eff}$ as a function of load as calculated from the first derivative of the curve in Figure \ref{figure2}a. We fit an offset power law function (i.e., $f(x) = \mathrm{a}x^{\mathrm{b}} +\mathrm{c}$) to these data to allow $R_\mathrm{eff}$ to be easily estimated for any given load. We only fit this function to data comparing fused silica and sapphire because the data involving glassy carbon correspond to smaller loads (Figure \ref{figure1}a), for which the data are considerably noisier. The data trend in Figure \ref{figure2}b matches the expectation of convergence to a single value for $R_\mathrm{eff}$ at larger loads. Comparing Figure \ref{figure6} with Figure \ref{figure2}b, we note that at higher loads, the calculations following \citet{li2013effects} overestimate the values for $R_\mathrm{eff}$, which could lead to underestimations of stress \citep[e.g.,][]{bei2016tale, gao2016strength}(see Table \ref{table_resultsSummary}).

\subsubsection{Determination of machine stiffness}
\label{Smach_calculation}

\begin{figure*}
  \includegraphics{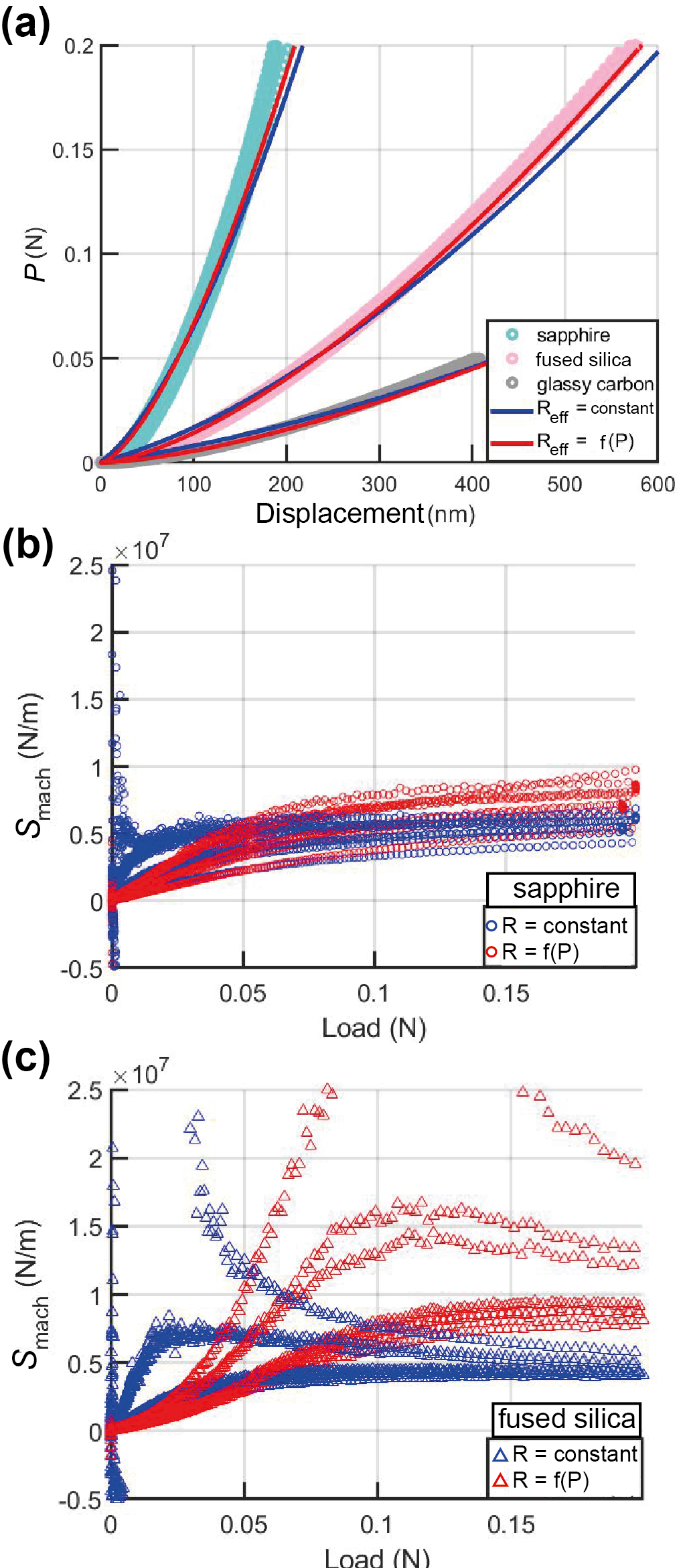}
\caption{a) Load-displacement data for a spherical tip with a nominal radius of 20 \textnormal{\mu}m in different materials with known moduli overlapped with predictions using a constant effective radius (blue) and radius as a function of depth (red) in Equation \ref{eqn:h_e}. b) Calculations of machine stiffness using sapphire data in Equation \ref{eqn:def_h_rep2}. c) Calculations of machine stiffness using fused silica data in Equation \ref{eqn:def_h_rep2}.}
\label{figure3-S-mach}       
\end{figure*}

For a robust interpretation of stress-strain curves obtained using spherical indentation, the impact of machine stiffness has to be determined for each combination of instrument and tip and can be reported as a function of load \citep{li2013effects, kang2012, cabibbo2012international}. To assess the magnitude of the machine stiffness, Figure \ref{figure3-S-mach}a compares the reported loads and displacements to the predicted displacement using Equation \ref{eqn:h_e}. In this equation, we implement $R_\mathrm{eff}$ either as a constant (blue curve, similar to Figure \ref{figure2}a) or as a function of load (red curve, similar to Figure \ref{figure2}b). Although the load-dependent $R_\mathrm{eff}$ is in better agreement with the reported data, there is still some mismatch in the displacements, which corresponds to $h_\mathrm{err}$ and which we attribute to the stiffness of the machine-tip pair. We use $R_\mathrm{eff}$ as a function of load in conjunction with $h_{\mathrm{err}}$ and the known value of $S_\mathrm{default}$, stated in Section \ref{expMethod}, to calculate $S_\mathrm{mach}$ according to Equation \ref{eqn:def_h_err}.  We plot this machine stiffness, $S_\mathrm{mach}$, as a function of load for sapphire and fused silica in Figures \ref{figure3-S-mach}b and c. For comparison, we also plot $S_\mathrm{mach}$ assuming $R_\mathrm{eff}$ is constant. In Figures \ref{figure3-S-mach}b and c, we plot the result of calculations of $S_\mathrm{mach}$ amongst tests in sapphire and fused silica, as there is no particular reason for using one data set over another when substituting in Equation \ref{eqn:def_h_err}. We emphasize that the values of $S_\mathrm{mach}$ calculated with for sapphire and fused silica both converge to $\sim 0.7*10^7$ N/m for a tip with a nominal radius of 20 \textnormal{\mu}m (Figure \ref{figure3-S-mach}). Note that the variability in $S_\mathrm{mach}$ resulting from tests on sapphire (Figure \ref{figure3-S-mach}b) and fused silica (Figure \ref{figure3-S-mach}c) is directly related to how well the Hertzian prediction in Figure \ref{figure3-S-mach}a fits the reported data. As demonstrated in Figure \ref{figure3-S-mach}a, the elastic prediction overestimates the reported displacement, which means $h_\mathrm{rep}-h_\mathrm{e}$ is negative, and therefore, the resulting effective machine-tip stiffness ($P/(h_\mathrm{rep}-h_\mathrm{e})$) is negative. This effective negative effective machine-tip stiffness does not imply that the machine actually has negative stiffness. Instead, the negative effective stiffness results from the default stiffness used by the software, $S_\mathrm{default}$, being smaller than the actual machine-tip stiffness. 

The stiffness of the machine-tip pair also needs to be accounted for when evaluating the reported harmonic contact stiffness. We model the machine-tip pair and the sample response as a series of elastic elements \citep{li2002CSMreview} and propose a correction for the reported harmonic contact stiffness using
\begin{equation}
\frac{1}{S_\mathrm{corr}} = \frac{1}{S_\mathrm{rep}} +  \frac{1}{S_\mathrm{default}} - \frac{1}{S_\mathrm{mach}},
\label{eqn:S_corr_def}
\end{equation}
where $S_\mathrm{corr}$ is the corrected contact stiffness and $S_\mathrm{rep}$ is the reported contact stiffness collected with $S_\mathrm{default}$. We implement Equation \ref{eqn:S_corr_def} with a variable $S_\mathrm{mach}$ by linearly fitting $S_\mathrm{mach}$ as a function of load in Figure \ref{figure3-S-mach}b. We recognize that these data depart from linearity but implement a linear fit as a practical approximation in line with previous studies \citep[e.g.,][]{li2013effects}. In this step, choices regarding the range of data used in the fitting will impact the slope of the fitted line. This is the key step in which we recommend an iteration between the fitting procedure and the outcome of the data correction verified in the benchmarks presented in the next section. 

\begin{figure*}
  \includegraphics[width=0.9\textwidth]{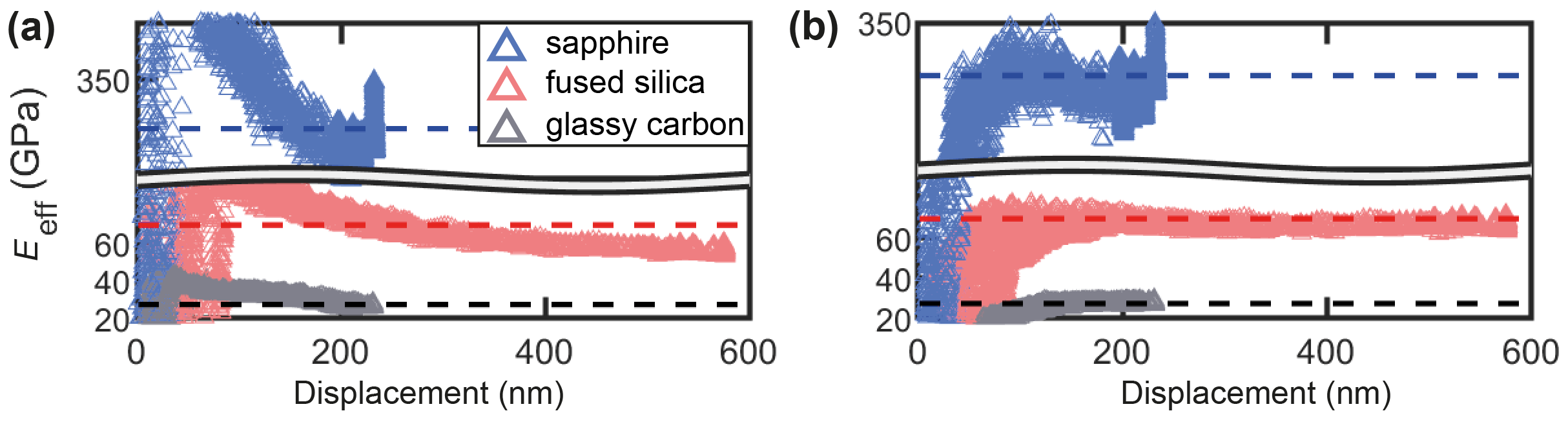}
\caption{Comparison of Young's modulus calculated using Equation \ref{eqn:E_def2} to expected values. Young's modulus is calculated using either a) a constant effective radius and reported values for the harmonic contact stiffness or b) a load-dependent effective radius and contact stiffness corrected according to Equation \ref{eqn:S_corr_def}. Note the differences in vertical scales. The reference values of the effective Young's modulus are plotted with dashed lines, and are calculated using the values in Table \ref{table_experiemntSummary} and Equation \ref{eqn:E_eff_def}. These tests were conducted with a tip with nominal radius of $R_\mathrm{n}= 5$ \textnormal{\mu}m.}
\label{figure4-modulus}       
\end{figure*}

\subsubsection{Benchmarks}
\label{Benchmarks}

Calibrations of spherical tips can be assessed with a variety of key benchmarks. The most common benchmark in nanoindentation is observation of a constant Young's modulus with depth \citep[e.g.,][]{leitner2018essential, phani2021measurement}. Following the derivation from \citet{hackett2019evaluation} and using Equations \ref{eqn:h_e} and \ref{eqn:S_corr_def}, we can express the effective Young's modulus as
\begin{equation}
E^*_\mathrm{eff} = \sqrt{\frac{S_\mathrm{corr}^3}{6PR_\mathrm{eff}}}. 
\label{eqn:E_def2}
\end{equation}

Using Equation \ref{eqn:E_def2}, we assess the impact of different corrections and formulations for $R_\mathrm{eff}$ on the Young's modulus. Figure \ref{figure4-modulus}a demonstrates that using a constant effective radius and the reported continuous stiffness measurement results in significant variability of the Young's modulus with depth. In contrast, Figure \ref{figure4-modulus}b demonstrates that using a load-dependent $R_\mathrm{eff}$ (e.g., Figure \ref{figure2}b) results in the convergence of $E_\mathrm{eff}$ towards expected values at relatively shallow depths. 

Another important benchmark is a consistent effective tip shape measured in multiple reference materials over the elastic depth range. A plot of the contact depth, $h_\mathrm{c}$, versus the contact radius, $a$, determined from the harmonic stiffness should essentially represent a profile of the effective indenter shape \citep{merle2012experimental, pathak2015sphericalreview} (Figure \ref{fig:Supplementary-cartoon}). In the case of elastic spherical indentation, the effective tip shape is expected to be the same during loading and unloading for a given tip regardless of which reference material is used. We calculate the contact depth (see Figure \ref{fig:Supplementary-cartoon}), $h_\mathrm{c}$, using \citep{sneddon1965, oliverandpharr92, pathak2015sphericalreview}
\begin{equation}
h_\mathrm{c} = h_\mathrm{rep} - \xi \frac{P}{S}, 
\label{eqn:def_h_c}
\end{equation}
where $\xi$ is a geometric factor equal to 0.75 for spherical indents \citep{fischer2011analysis, oliverandpharr92}. Implementing corrections associated with the system stiffness, we calculate the corrected contact depth,  $h^*_\mathrm{c}$, as
\begin{equation}
h^*_\mathrm{c} = h_\mathrm{rep} - h_\mathrm{err} - \xi \frac{P}{S_\mathrm{corr}}.
\label{eqn:h_c_corr}
\end{equation}
We calculate the contact radius using both reported and corrected values of the contact stiffness, \citep{oliverandpharr92, pathak2015sphericalreview}
\begin{gather}
a = \frac{S_\mathrm{rep}}{2E_\mathrm{eff}} \\
\nonumber \mathrm{and}\\
a_\mathrm{corr} = \frac{S_\mathrm{corr}}{2E_\mathrm{eff}}.
\label{def_a_corr}
\end{gather}
We compare the effective tip shape with the shape predicted for a perfect sphere with contact radius $a_\mathrm{pred}$given by
\begin{equation}
a_\mathrm{pred} = \sqrt{2R_\mathrm{i}h_\mathrm{c} - h_\mathrm{c}^2}.
\label{eqn:a_pred}
\end{equation}
Note that for perfectly elastic indents $h_\mathrm{c} = h_\mathrm{e}/2$ and $R_\mathrm{i} = R_\mathrm{eff}$ \citep{basu2006determination, pathak2015sphericalreview}.  

\begin{figure*}
  \includegraphics[width=0.9\textwidth]{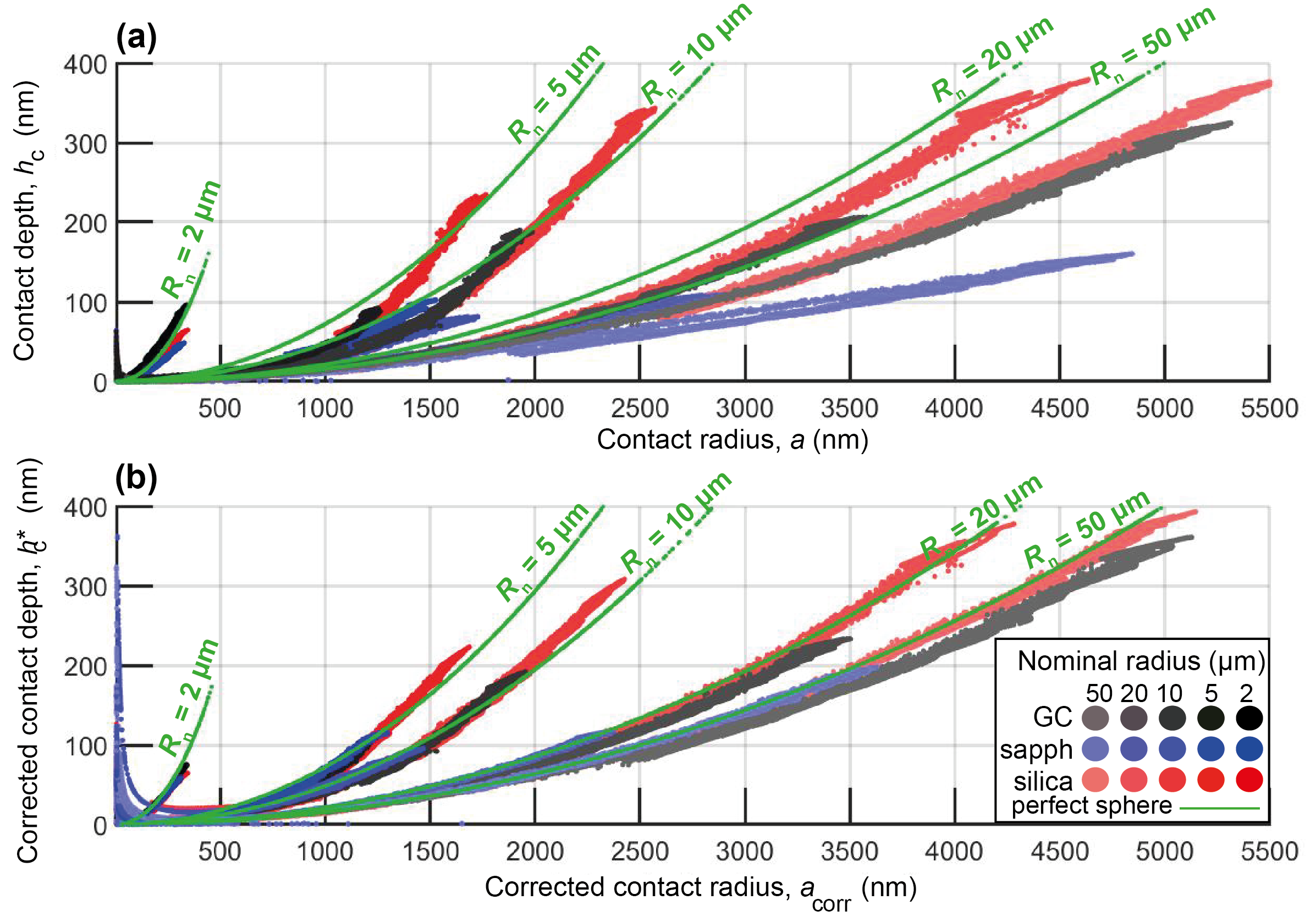}
\caption{a) The effective tip shape calculated with data as reported and b) the effective tip shape calculated with corrected displacement and harmonic stiffness data for spherical tips with load-dependent radii. The green curve represents the tip shape for a perfect sphere, with the contact radius calculated using Equation \ref{eqn:a_pred} with values for $R_{\mathrm{n}}$ from Figure \ref{figure6}.}
\label{figure5-tip-shape}       
\end{figure*}

Figure \ref{figure5-tip-shape} summarises the impact of the proposed calibration routines and data corrections on the effective tip shape. We observe in Figure \ref{figure5-tip-shape}a that the uncorrected data display significant discrepancies between the apparent tip shape and the ideal predicted tip shape calculated using the constant radii summarised in Figure \ref{figure6}. We also observe that, the apparent tip shape varies depending on the reference material, with the sapphire data set exhibiting the most significant discrepancies. Figure \ref{figure5-tip-shape}b indicates that our proposed strategies for data correction result in good agreement between the apparent tip shape and the tip shape calculated with the best-fit, constant radius. These corrections also result in a consistent apparent tip shape among different reference materials.

\subsection{Characterization of materials with unknown Young's moduli}
\label{applicationMaterials}
In section \ref{referenceMaterials} we determined that two key parameters (the effective radius and machine stiffness) are needed for each combination of machine and indenter tip (Figure \ref{figure5-tip-shape}) to calculate the Young's modulus and stress-strain curves using load-displacement data (Figure \ref{fig:Supplementary3-methods}, Supplementary Materials). In this section, we demonstrate how these calibrations may be implemented to extract stress-strain curves in samples with potentially unknown moduli, using W, Ti, and olivine as examples. The details of these experiments are described in Section \ref{expMethod} (see Table \ref{table_experiemntSummary}). 

\subsubsection{Zero-point correction}
\label{ZPC}
A key step in analysing reported load-displacement data is the identification of the point of zero-displacement and zero-load, i.e., the zero-point correction. The impact of the zero-point correction on the estimations of $R_\mathrm{eff}$ and stress-strain curves has been a focus in previous analyses of spherical nanoindentation data \citep{kalidindi2008ZPC, pathak2015sphericalreview,moseson2008ZPC}. In our approach for calibrating spherical tips using reference materials of known elastic moduli, the last term in Equation \ref{eqn:delta_h_rep} already accounts for errors in the point of initial contact. However, the methods discussed above do not provide a means for finding the effective point of zero contact in data acquired in materials with unknown elastic moduli. 

\citet{kalidindi2008ZPC} proposed a method for determining the effective point of contact by using the relationship between the reported load, displacement, and harmonic stiffness measurements for the elastic portion of an indentation test, 
\begin{equation}
S_\mathrm{rep} = \frac{3P}{2h_\mathrm{e}} = \frac{3({P} - P_\mathrm{0})}{2({h_\mathrm{e}} - h_\mathrm{0})},
\label{eqn:ZPC1}
\end{equation}
where $h_\mathrm{0}$ and $P_\mathrm{0}$ represent the values of displacement and load at the actual point of contact, respectively. These values can be found by re-arranging Equation \ref{eqn:ZPC1} to yield a linear relationship between ${P} - \frac{2}{3}S_\mathrm{rep}{h_\mathrm{e}}$ and $S_\mathrm{rep}$, with the slope equal to $- \frac{2}{3}h_\mathrm{0}$ and the vertical intercept equal to $P_\mathrm{0}$. Thus, the corrected load-displacement data are the reported data minus the values for $P_\mathrm{0}$ and $h_\mathrm{0}$. One major advantage of this approach is that no \emph{a priori} knowledge of $E_\mathrm{eff}$ or $R_\mathrm{eff}$ is necessary. This approach is also suitable for anisotropic materials \citep{kalidindi2008ZPC, pathak2015sphericalreview}. However, this method also relies on accurate identification of the elastic loading segment for fitting by linear regression, which is often a subjective procedure since, in practice, the transition between elastic and plastic deformation is not sharply defined \citep[e.g., Figure 4 in][]{kalidindi2008ZPC}. This difficulty in identifying the most appropriate segment of data for a linear fit introduces significant uncertainty in the calculated stress and strain \citep{weaver2016capturing}. 

In this section, we present an alternative formulation proposed by \citet{breithaupt2017}, which is adapted from \citet{kalidindi2008ZPC} for the zero-point correction of load-displacement data. We calculate the values for $P_\mathrm{0}$ and $h_\mathrm{0}$ by minimising the residual, $r^*$, between the data and predictions of perfect elasticity in the stress-strain curve. Thus, we define the residual, $r$, as \citep{breithaupt2017}
\begin{equation}
r = \sigma - E_\mathrm{eff}\varepsilon,
\label{eqn:residuals}
\end{equation}
where the indentation stress, $\sigma$, and strain, $\varepsilon$, are defined according to \citet{kalidindi2008ZPC} and \citet{pathak2015sphericalreview} as
\begin{equation}
\sigma = E_\mathrm{eff}\varepsilon,
\label{eqn:stress-strain}
\end{equation}
\begin{equation}
\sigma = \frac{P}{\pi a^2},
\label{eqn:stress}
\end{equation}
\begin{equation}
\varepsilon = \frac{4h_\mathrm{e}}{3 \pi a}.
\label{eqn:strain}
\end{equation}
We can substitute these definitions for stress, strain, and contact radius in Equation \ref{eqn:residuals} and rearrange to yield
\begin{equation}
r = \frac{4E_\mathrm{eff}^2}{3\pi} \Big(\frac{3P - 2S_\mathrm{rep}h_\mathrm{e}}{S_\mathrm{rep}^2}\Big).
\label{eqn:residual2}
\end{equation}
Recasting Equation \ref{eqn:residual2} in terms of reported and corrected values for load and displacement (similar to Equation \ref{eqn:ZPC1}) and summing the absolute error leads to the proportionality \citep{breithaupt2017}
\begin{equation}
r_\mathrm{total} \propto \sum \Bigg|\Bigg| \frac{3({P} - P_\mathrm{0}) - 2S_\mathrm{rep}({h_\mathrm{e}} - h_\mathrm{0})}{S_\mathrm{rep}^2} \Bigg|\Bigg|.
\label{eqn:residual-tot}
\end{equation}

The residual in Equation \ref{eqn:residual-tot} describes the departure from elasticity. In practice, we find the values for $h_\mathrm{0}$ and $P_\mathrm{0}$ by minimising $r_\mathrm{total}$. We subtract the values for $h_\mathrm{0}$ and $P_\mathrm{0}$ from the uncorrected reported values to yield the corrected data for the point of zero contact. The advantage of this approach is that a significantly larger portion of the data set is used than in the linear regression by \citet{kalidindi2008ZPC}, and therefore the correction is less sensitive to accurate identification of the elastic segment.

\subsubsection{Calculation of Young's moduli}
\begin{figure*}
  \includegraphics[width=0.9\textwidth]{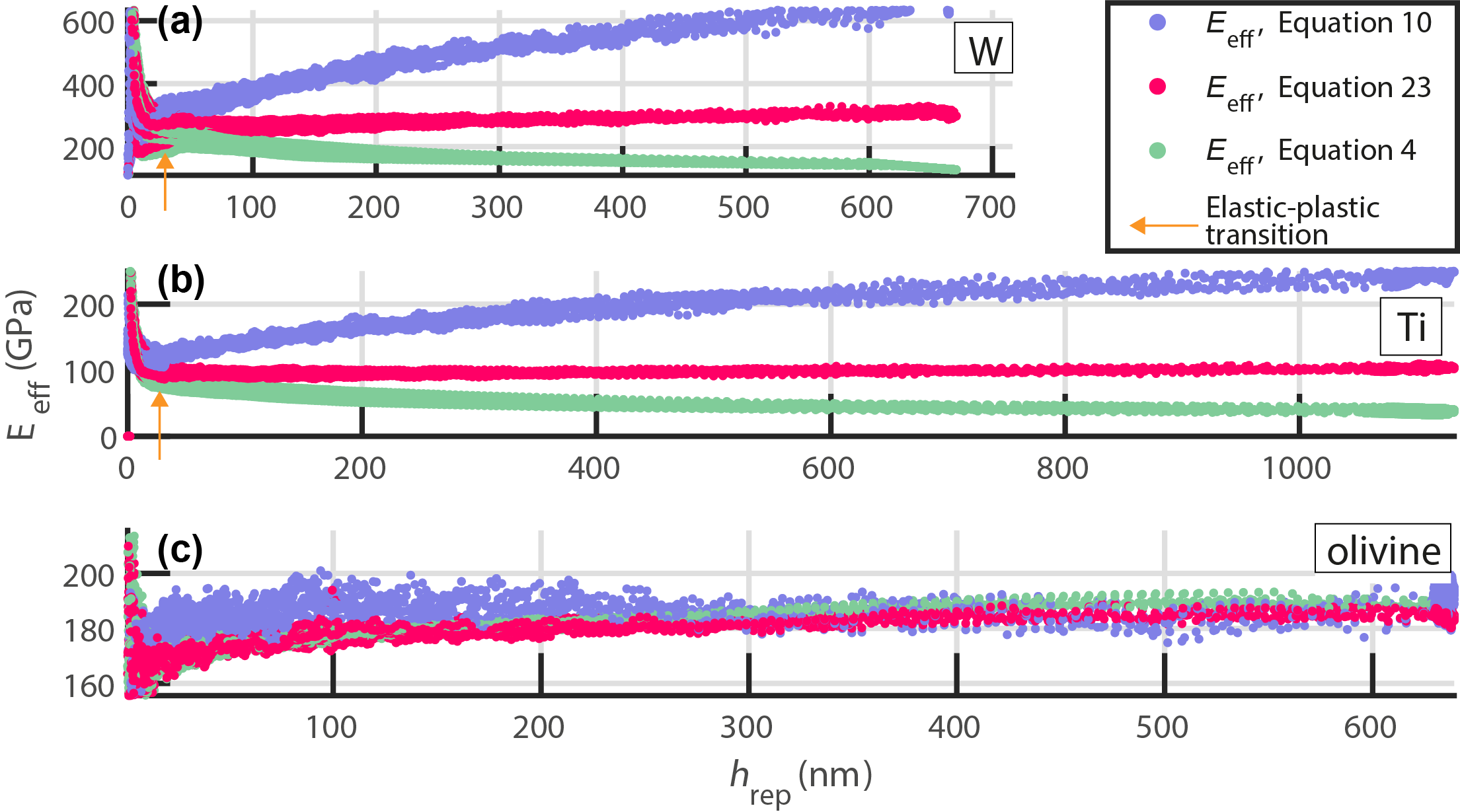}
\caption{Measurements of Young's modulus for a) tungsten (average $E_\mathrm{eff} = 306.7 \pm 14.2$ GPa), b) titanium (average $E_\mathrm{eff} = 99.5 \pm 4.7$ GPa), and c) olivine (average $E_\mathrm{eff} = 183.8 \pm 8.8$ GPa) using data collected with a tip with a nominal radius of 50 \textnormal{\mu}m. The orange arrow indicates the approximate position of the elastic-plastic transition, which is absent in the fully elastic experiments on olivine. The effective Young's modulus is calculated using Equations \ref{eqn:h_e} (green), \ref{eqn:E_eff_leitner} (red), and \ref{eqn:E_def2} (purple) to demonstrate that equations assuming elastic-only experiments diverge after the elastic-plastic transition.}
\label{figure7}       
\end{figure*}

After the reported load-displacement data (${P}$ and ${h}$) are corrected for the effective point of contact by subtracting $P_\mathrm{0}$ and $h_\mathrm{0}$, respectively, we calculate the effective elastic modulus. In stiffer materials (e.g., olivine) with a clearly identifiable segment of elastic load-displacement, $E_\mathrm{eff}$ can be determined from Equation \ref{eqn:h_e}, which requires the elastic displacement and effective radius to be known. However, any error in $R_\mathrm{eff}$ or in determining the elastic segment will significantly impact estimations of the effective Young's modulus. Importantly, if plastic yield occurs, then the elastic displacement is not explicitly known and the sample curvature, $R_\mathrm{s}$, becomes finite and modifies $R_\mathrm{eff}$ according to Equation \ref{eqn:R_eff_def}. This influence of plastic yielding is not easily accounted for and leads to an erroneous depth dependence of $E_\mathrm{eff}$.  

We can also use Equation \ref{eqn:E_def2} to evaluate the Young's modulus over the elastic segment, as previously done for reference materials in Figure \ref{figure4-modulus}. However, Equation \ref{eqn:E_def2} also relies on $R_\mathrm{eff}$ regardless of whether or not we use a constant radius or a load-dependent radius and is therefore subject to the same effects as the previous method if plastic yielding occurs. 

Finally, we can calculate $E_\mathrm{eff}$ by assuming a constant radius and using Equations \ref{def_a_corr} and \ref{eqn:a_pred},
\begin{equation}
E_\mathrm{eff} = \frac{\sqrt{\pi}}{2}\frac{S_\mathrm{corr}}{\sqrt{2\pi R_\mathrm{i}h_\mathrm{c} - \pi h_\mathrm{c}^{2}}}.
\label{eqn:E_eff_leitner}
\end{equation}

Because these equations are inherently based on the geometry of an ideal spherical tip, they are dependent on $R_\mathrm{i}$ and not on $R_\mathrm{eff}$. Therefore, this method is not influenced by plastic yield of the surface \citep[e.g.,][]{leitner2018essential}. We implement Equation \ref{eqn:E_eff_leitner} with corrected values for the harmonic stiffness measurement and the contact depth. As previously illustrated in Figure \ref{figure5-tip-shape}, these corrections result in load-displacement data that can be described with a constant $R_\mathrm{eff}$. We implement Equation \ref{eqn:E_eff_leitner} using constant values for $R_\mathrm{eff}$ (Table \ref{table_resultsSummary}, column 1) to compute $R_\mathrm{i}$. This approach is complementary to the one presented by \citet{leitner2018essential}, who modify the parameters in Equation \ref{eqn:a_pred} describing the geometry of the perfect tip for a given material. The advantage of our approach is that it inherently accounts for the effects of the machine stiffness. That this approach is still applicable after plastic yield is particularly advantageous for materials with a short or noisy elastic segment. Moreover, any issues related to tip calibrations or machine stiffness corrections will result in moduli that are not constant with indentation depth, flagging if there are issues with calibration. 

Figure \ref{figure7} displays examples of each these three approaches to measuring Young's modulus using tests in tungsten, titanium, and olivine conducted with a tip with a nominal radius of $R_\mathrm{n} = 50$ \textnormal{\mu}m. As expected, the calculated moduli are comparable for elastic indentation at small displacements but diverge after plasticity initiates, at which point Equations \ref{eqn:h_e} and \ref{eqn:E_def2} are no longer applicable. Moreover, Equation \ref{eqn:E_eff_leitner} implemented with corrected values for $h_\mathrm{c}$ and $R_\mathrm{i}$ results in essentially constant values for Young's modulus, even after yield, that are in agreement with previously published values for the tested materials \citep[e.g,][]{kumamoto2017size, pathak2015sphericalreview}.  

\subsubsection{Calculation of stress-strain curves and the yield point} 

\begin{figure*}
  \includegraphics[width=0.99\textwidth]{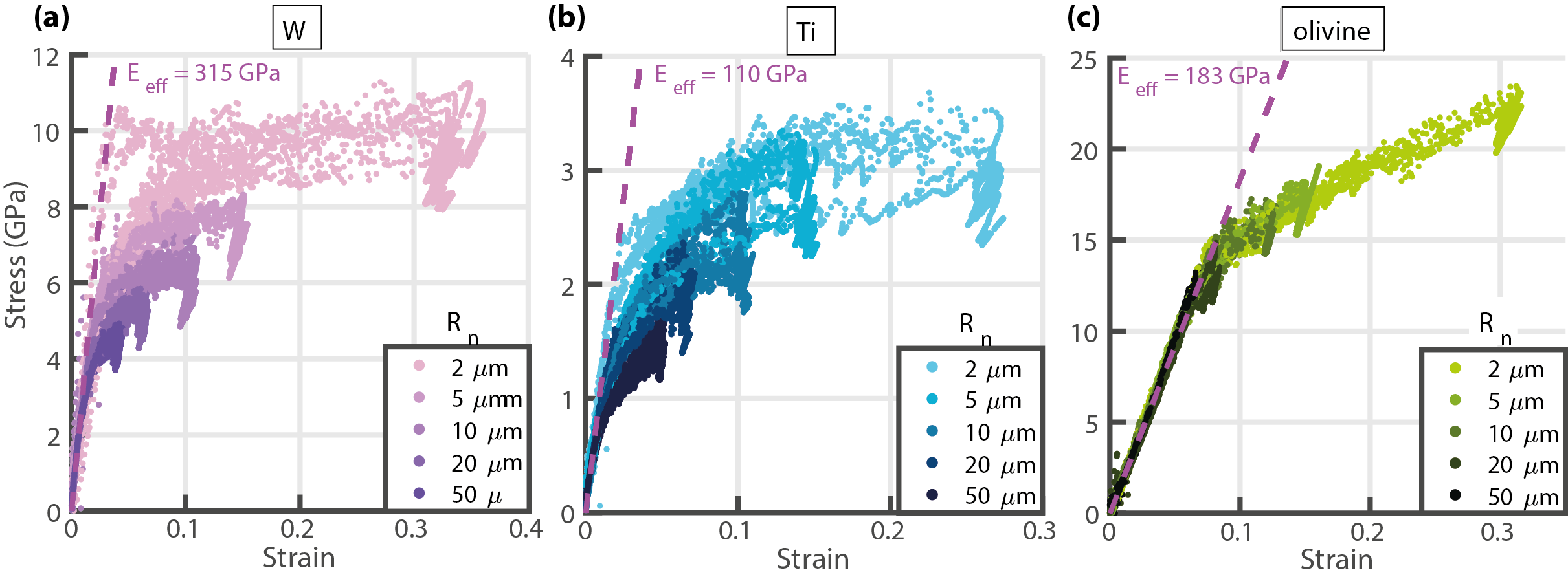}
\caption{Summary of stress-strain curves in a) tungsten, b) titanium, and c) olivine collected with spherical tips of varying radii. The dashed lines have the slope equal to the average $E_\mathrm{eff}$ across all indents. The displayed values of $E_\mathrm{eff}$ correspond to  $E_\mathrm{s} = 400 \pm 9$ GPa in tungsten, $E_\mathrm{s} = 108 \pm 8$ GPa in titanium, and $E_\mathrm{s} = 205 \pm 9$ GPa in olivine. The corresponding load-displacement curves are presented in Supplementary Materials (Figure \ref{fig:Supplementary3-methods}).}
\label{figure8}     
\end{figure*}

Figure \ref{figure8} presents stress-strain curves calculated for tungsten, titanium, and olivine using the spherical tips summarised in Table  \ref{table_experiemntSummary} and Equations \ref{eqn:stress} and \ref{eqn:strain} for stress and strain, respectively, as defined by \citet{pathak2015sphericalreview} and \citet{kalidindi2008ZPC}. All materials display an indentation size effect, in which the yield stress increases with decreasing effective radius. Note that the polycrystalline materials display a greater variability of stress values than the olivine single crystal, due to plastic anisotropy \citep[e.g.,][]{britton2010effect, weaver2016capturing}. Figures \ref{figure8}a and b display $E_\mathrm{eff}$ values corresponding to averages of $E_\mathrm{s} = 400 \pm 9$ GPa in W, and $E_\mathrm{s} = 108 \pm 8$ GPa in Ti, which is in line with published values for W and on the lower end of the spectrum for Ti \citep{pathak2015sphericalreview, britton2010effect, weaver2016capturing}. We calculate an indentation flow stress ranging between 4 and 10 GPa for W. In comparison, studies of indentation size-effects using sharp tips document a hardness of 4 GPa in W at the greatest contact depths \citep[e.g.,][]{javaid2018indentation, maier2015microstructure}. We calculate an indentation flow stress ranging between 1 and 3.5 GPa for Ti, in agreement with published spherical nanoindentation results in single crystals \citep{weaver2016capturing}. We note the absence of pop-ins in the indentation stress-strain curves collected in these polycrystalline samples, which is in contrast to the presence of pop-ins documented using spherical nanoindentation in both W \citep{patel2020spherical} and Ti single crystals \citep{weaver2016capturing}.

\begin{figure*}
  \includegraphics[width=0.99\textwidth]{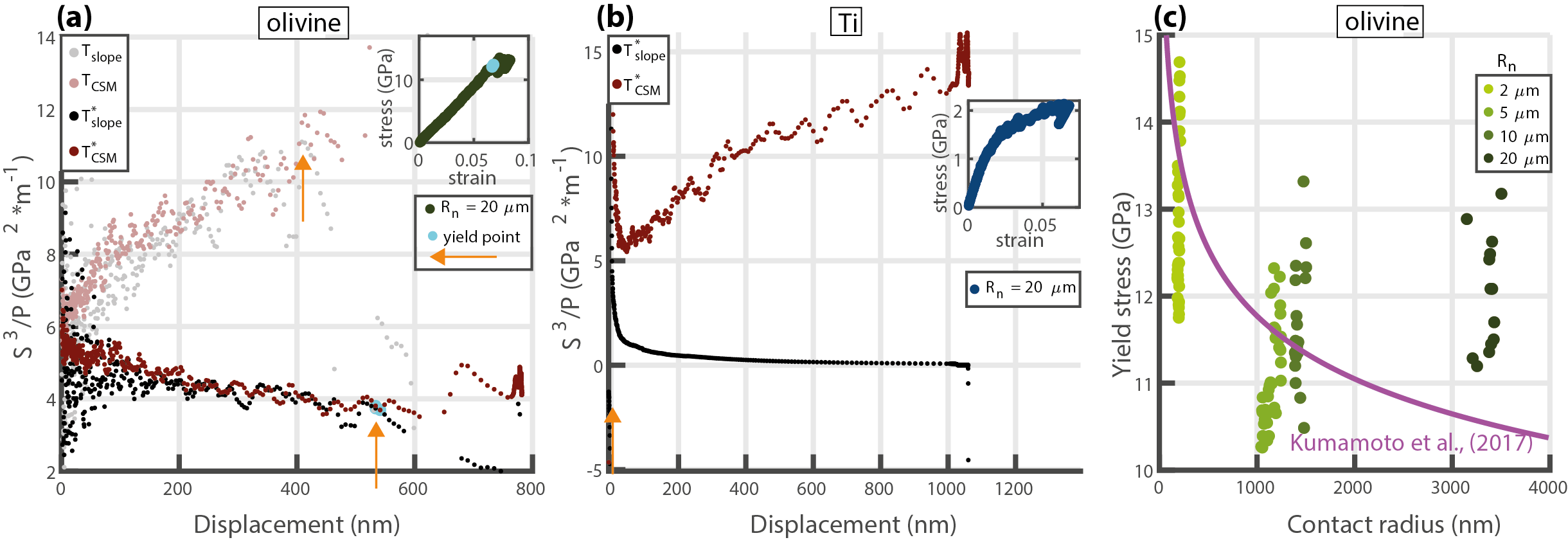}
\caption{a) Example of $S^3/P$ parameter calculations using Equations \ref{eqn:S_slope} and \ref{eqn:S_CSM} proposed by \citet{hackett2019evaluation} using both reported (light colours, notes as $T_\mathrm{slope}$ and $T_\mathrm{CSM}$ ) and corrected (dark colours, notes as $T^*_\mathrm{slope}$ and $T^*_\mathrm{CSM}$) values. The corresponding stress-strain curve is presented in the insert. A filled-blue circle and an orange arrow mark the yield point on the corrected data in a) and the insert. The data were collected with a tip with $R_{n} = 20$ \textnormal{\mu}m. A smoothing window with an interval of 5 is applied to $T_\mathrm{slope}$ and $T^*_\mathrm{slope}$. We present calculations for a) olivine and b) titanium. The elastic-plastic transition in b) is indistinguishable as it occurs at small indentation depths. c) Yield stress corresponding to the divergence point in a) for olivine across tips with varying radii. The power law fit with an exponent of -0.09 has been determined by \citet{kumamoto2017size} using spherical nanoindentation on similar samples, but deploying a different strategy for defining the yield stress.}
\label{figure9}     
\end{figure*}

Furthermore, Figure \ref{figure7} highlights that the divergence of the moduli calculated with different formulations marks the yield point of the material. This result has a similar basis to the method proposed by \citet{hackett2019evaluation} for identifying the yield point. Their method, however, does not require \emph{a priori} knowledge of $R_\mathrm{eff}$ or $E_\mathrm{eff}$. To achieve this, \citet{hackett2019evaluation} compute the value $S^3/P$ using two different methods for computing the contact stiffness. In the first method, the contact stiffness is obtained by differentiation \citep{diffxy} of the load-displacement curve, such that \citep{hackett2019evaluation} 
\begin{equation}
T_\mathrm{slope} = \big(\frac{dP}{dh}\big)^3P^{-1}.
\label{eqn:S_slope}
\end{equation}
This method calculating the stiffness is only valid for perfectly elastic portions of the loading curves. The second method uses the reported harmonic contact stiffness \citep{hackett2019evaluation}
\begin{equation}
T_\mathrm{CSM} = \frac{S^3}{P},
\label{eqn:S_CSM}
\end{equation}
which is valid even if there is some plastic strain. These two estimates of $S^3/P$ will be equivalent for elastic deformation and will diverge after plastic yield. Importantly, the first method relies on measurements of displacement and the second method relies on measurements of the harmonic contact stiffness, both of which will depend strongly on the corrections described above.

In Figure \ref{figure9}a, we compute the elastic parameter $S^3/P$ using both of these methods to identify the yield point for an indent in olivine collected using a tip with a nominal radius of 20 \textnormal{\mu}m. We compare the yield point identified using reported values for $S$ and $h$ to the yield point identified using corrected values according to Equations \ref{eqn:def_h_err} and \ref{eqn:S_corr_def}. The light coloured data represents reported values, which are significantly different to the corrected data displayed in darker colours. The orange arrows and the blue filled symbols mark the divergence point in the two independent calculations of the $S^3/P$ parameter and the corresponding position of this point on the stress-strain curve for the same indent (insert in  Figure \ref{figure9}a). Figure \ref{figure9}a highlights that this method is highly sensitivity to the machine stiffness corrections and the zero-point correction, with the point of divergence shifted by about 150 nm after applying the corrections.

In contrast to olivine, we are unable to distinguish an elastic segment implementing the method proposed by \citet{hackett2019evaluation} in titanium and tungsten with these indentation contact sizes, as indicated by the immediately diverging values for $E_{\mathrm{eff}}$ at small depths in Figure \ref{figure7}. For example, Figure \ref{figure9}b displays the $S^3/P$ parameter computed for Ti using data collected with a tip with a nominal radius of 20 \textnormal{\mu}m, and corrected $S$ and $h$ values. The two calculations of the $S^3/P$ parameter diverge at very small indentation depths ($< 50$ nm) consistent with the onset of plasticity at small displacements. Thus, this method is suitable only for materials displaying an elastic segment of minimum $\approx$ 100 nm \citep{hackett2019evaluation}. Figure \ref{figure9} emphasizes that in ductile materials the zero-point correction and determination of the Young's modulus following the methods proposed by \citet{kalidindi2008ZPC} and \citet{pathak2015sphericalreview} rely on data collected at small depths, which is heavily impacted by the quality of the initial contact. 

To date, previous studies have implemented different conventions for determining the yield stress in spherical nanoindentation stress-strain curves \citep[e.g.,][]{herbert2001measurement}. In this study, we picked the point of divergence exemplified in Figure \ref{figure9}a for all indents in olivine and summarised the corresponding values in Figure \ref{figure9}c as a function of their contact radii. \citet{kumamoto2017size} used spherical nanoindentation and quantified a size effect in olivine in which the yield stress is proportional to the contact radius according to a power law with an exponent of $-0.09$. Figure \ref{figure9}c displays a size effect with some deviations from this power law. The discrepancy is likely due to the inherent differences in the definitions of the yield point, and possibly due to material anisotropy. \citet{kumamoto2017size} present stress-strain curves in annealed single crystals with a pop-in associated with dislocation nucleation and glide after a longer segment of elastic loading compared to deformed samples. This phenomenon, documented in materials with scarce dislocation sources, is attributed to the requirement of a larger deformation volume for activating dislocation sources \citep[e.g.,][]{shim2008different, morris2011popin, bei2016tale}. The method presented in Figure \ref{figure9} implements a definition of the yield stress as the end of the elastic loading segment \citep{hackett2019evaluation}. However, in the indents displaying a pop-in, the point of divergence in Figure \ref{figure9}a corresponds to stresses required for initiation of plasticity that are elevated relative to typical yield stresses because of the lack of dislocation sources. For example, \citet{kumamoto2017size} define yield stress as the intersection of the projected slope of the hardening curve at high strains with the elastic-loading curve, in line with other studies using spherical nanoindentation \citep[e.g.,][]{pathak2015sphericalreview}. Therefore, the apparent overestimate of the stresses in Figure \ref{figure9}c for the largest contacts is consistent with the differences in the two conventions for the yield stress. Other studies implement a definition of the yield stress as the stress at a strain chosen by convention for a meaningful comparison with data obtained in uniaxial macroscale tests \citep[e.g., 0.2\% in][]{patel2016correlation, iskakov2022multiresolution, courtright2021critical, khosravani2018multiresolution, weaver2016capturing}. The calculated yield stress is a key outcome of spherical indentation and the choice of convention can influence the quantification of size effects in materials \citep[e.g.,][]{herbert2001measurement}. 

\section{Conclusion}
\label{conclusion}

We investigated in detail and further improved published methodologies for independently determining two key instrument parameters in spherical nanoindentation: system stiffness and effective radius of the indenter tip. To this end, we collect elastic data in reference materials with known moduli and highlight complexities in determining the effective radius using published protocols \citep[e.g.,][]{li2013effects}. We suggest a routine underpinned by parameterizing the effective radius as a function of load to overcome experimental errors. We benchmark our methodology against key criteria in spherical nanoindentation. We implement the tip calibrations on data collected on materials with unknown Young's modulus and varying ductility to calculate stress-strain curves. A summary diagram of this routine can be found in Supplementary Materials (Figure \ref{fig:Supplementary2-methods}). These curves reveal a spherical indentation size effect in which the stress increases with indentation depth and with decreasing contact radius. We also test the influence of instrument parameters on published methods for determining the yield stress in materials with a significant elastic loading segment \citep[e.g.,][]{hackett2019evaluation} and highlight the importance of consistency when establishing a convention for determining the yield stress. These improvements of the spherical nanoindentation technique are critical for refining the measurement and corrected values of the contact stiffness, given the importance of this measurement to the extraction of stress-strain curves, calculation of Young's modulus, and analysis of pop-ins \citep[e.g., in Berkovich nanoindentation][]{phani2021measurement}.

\section{Experimental aspects}
\label{expMethod}

Nanoindentation tests with spherical tips and continuous stiffness measurements were performed with a displacement-controlled indenter (Nanoindenter G200, Agilent Technologies) with the frequency target set at 45 Hz, the harmonic displacement target set at 2 nm, and the loading rate divided by load set at $\frac{\dot P}{P}=0.05 s^{-1}$. We used diamond spherical tips ($E_\mathrm{i} = 1141$ GPa, $v_\mathrm{i} = 0.07$) with a 2--50 \textnormal{\mu}m range of nominal tip radii (Table \ref{table_experiemntSummary}). The data collection was undertaken with the default machine stiffness, $S_\mathrm{default} = 3.67*10^6$ N/m.

To assess tip calibration and data analysis routines, we mounted reference materials with known moduli (fused silica, sapphire, and glassy carbon) on the same stub as materials with unknown elastic moduli (olivine, W, Ti) using the smallest amount of epoxy necessary. This set-up mediates differences in assembly stiffness due to the mounting substrate. The olivine single crystal sample used in this study is the undeformed sample also used by \citet[sample MN1,][]{kumamoto2017size, wallis2020dislocation}. Both the titanium and tungsten samples are undeformed, commercially-available pure samples, with grain sizes in the ranges of 10--50 \textnormal{\mu}m and 10--100 \textnormal{\mu}m, respectively. 

We performed 270 nanoindentation tests in total, with at least 9 tests for each tip reported in Table \ref{table_experiemntSummary}.
Nanoindentation experiments in materials with known Young’s modulus were performed at small loads, resulting in elastic load-displacement curves (Table \ref{table_experiemntSummary}). We measured the harmonic contact stiffness throughout the loading and unloading paths in all experiments.  

\begin{table*}[ht]
\centering
\caption{Summary of experiments. Note that the experiments in the reference materials with known elastic moduli are experiments in the elastic deformation regime only.}
\label{table_experiemntSummary}       
\begin{tabular}{llllllllll}
\hline\noalign{\smallskip}
{} & \multicolumn{5}{|c|}{Nominal tip radius, $R_\mathrm{n}$} & {} & {} \\ 
{} & 2 \textnormal{\mu}m  & 5 \textnormal{\mu}m & 10 \textnormal{\mu}m & 20 \textnormal{\mu}m & 50 \textnormal{\mu}m & {} & {}\\
Material & \multicolumn{5}{|c|}{Maximum load (N)} & Young's Modulus, $E_\mathrm{s}$, (GPa) & Poisson ratio, $\upsilon_\mathrm{s}$ & Grain size (\textnormal{\mu}m) \\ 

\noalign{\smallskip}\hline\noalign{\smallskip}
fused silica & $2.5*10^{-3}$ & $8*10^{-2}$&$1*10^{-1}$& $2*10^{-1}$&$3*10^{-1}$&72&0.17&single crystal\\
sapphire & $2.5*10^{-3}$ & $8*10^{-2}$&$1*10^{
-1}$&$2*10^{-1}$&$3*10^{-1}$ & 420&0.28&single crystal\\
glassy carbon & $1.5*10^{-3}$ & $1*10^{-2}$&$3*10^{-2}$&$5*10^{-2}$&$1*10^{-1}$ & 34&0.27&-\\
olivine & $8*10^{-2}$ & $2.5*10^{-1}$ &$3.8*10^{-1}$&$6.5*10^{-1}$&$6.5*10^{-1}$&-&0.24&single crystal\\
Ti & $2.5*10^{-2}$ & $8*10^{-2}$ & $1.5*10^{-1}$ & $3*10^{-1}$ & $4*10^{-1}$ & - & 0.35 & 35\\
W & $5.5*10^{-2}$ & $2*10^{-1}$ &$3.2*10^{-1}$&$6*10^{-1}$&$6*10^{-1}$&-&0.29&$>50$\\
\noalign{\smallskip}\hline
\end{tabular}
\end{table*}

\begin{acknowledgements}
This work has been stimulated by conversations in the Oxford Micromechanics Group meetings. DA acknowledges useful conversations and input from Kathryn Kumamoto, David Armstrong, and Thomas Breithaupt. 
\end{acknowledgements}

\section*{Declarations}
\subsection*{Funding}
DA is grateful to the UK National Environmental Research Council, and the Oxford Doctoral Training Partnership for DPhil studentship and funding from grant NE/L002612/1. AK acknowledges support from the UK Engineering and Physical Research Council under Fellowship grant EP/R030537/1. AJW and LH acknowledge support from the UK National Environmental Research Council under the grant NE/S00162X/1.  

\subsection*{Conflict of interest/Competing interests}
On behalf of all authors, the corresponding author states that there is no conflict of interest.

\subsection*{Ethics approval}
Not applicable
\subsection*{Consent to participate}
Not applicable
\subsection*{Consent for publication}
\subsection*{Availability of data and materials}
Nanoindentation data are available under an open access license, with the following DOI: 10.5281/zenodo.7607547

\subsection*{Code availability}
The authors will make the code available upon reasonable request.

\subsection*{Authors' contributions}
We report authors' contributions according to CRediT taxonomy. Conceptualization: [Anna Kareer], [Diana Avadanii], [Lars Hansen]; Investigation: [Anna Kareer], [Diana Avadanii]; Methodology: [Diana Avadanii], [Lars Hansen]; Formal analysis and investigation: [Diana Avadanii]; Writing - original draft preparation: [Diana Avadanii]; Writing - review and editing: [Anna Kareer], [Lars Hansen], [Angus Wilkinson]; Resources: [Anna Kareer], [Angus Wilkinson]; Supervision: [Lars Hansen], [Angus Wilkinson];

\bibliography{refs.bib}   

\clearpage
\appendix
\counterwithin{figure}{section}
\counterwithin{table}{section}
\section{Supplementary Materials}
\clearpage

\begin{figure*}
  \includegraphics[width=0.9\textwidth]{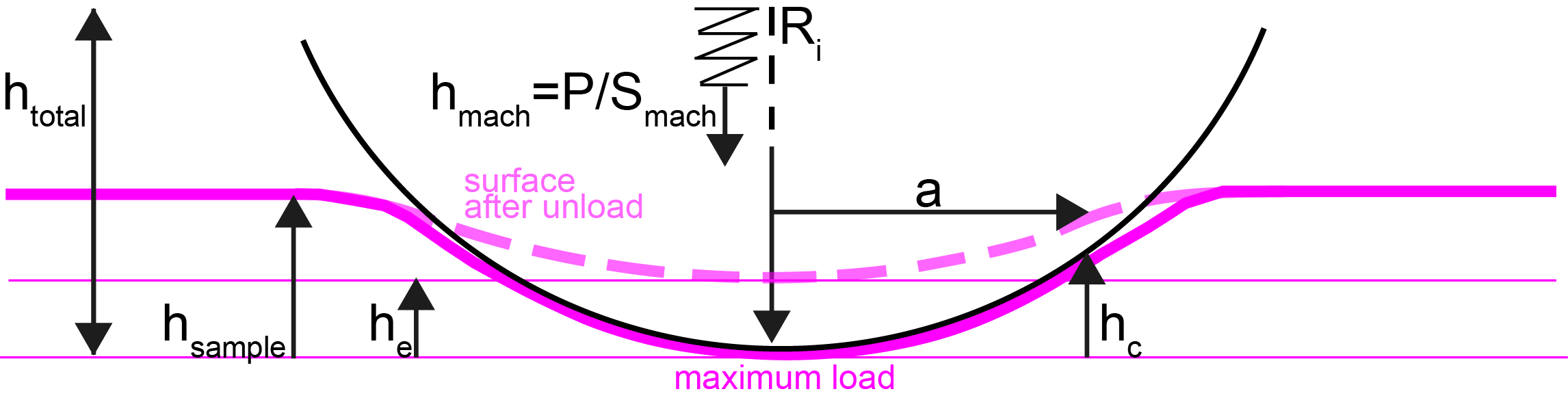}
\caption{Schematic illustration of a spherical indent and the displacement during nanoindentation, adapted after \citet{pathak2015sphericalreview} and \citet{li2013effects}. The total measured displacement, $h_{\mathrm{total}}$ is the summation of the displacement due to deformation of the sample, $h_{\mathrm{total}}$, and the deflection, $h_{\mathrm{mach}}$, induced by the combined stiffness from the indenter tip and the loading frame, $S_{\mathrm{mach}}$. The plot of contact depth, $h_{\mathrm{c}}$, calculated according to Equation \ref{eqn:def_h_c} against the contact area, $a$, calculated using Equation \ref{def_a_corr}, describes the effective tip shape during spherical nanoindentation.}
\label{fig:Supplementary-cartoon}     
\end{figure*}

\begin{figure*}
  \includegraphics[width=0.8\textwidth]{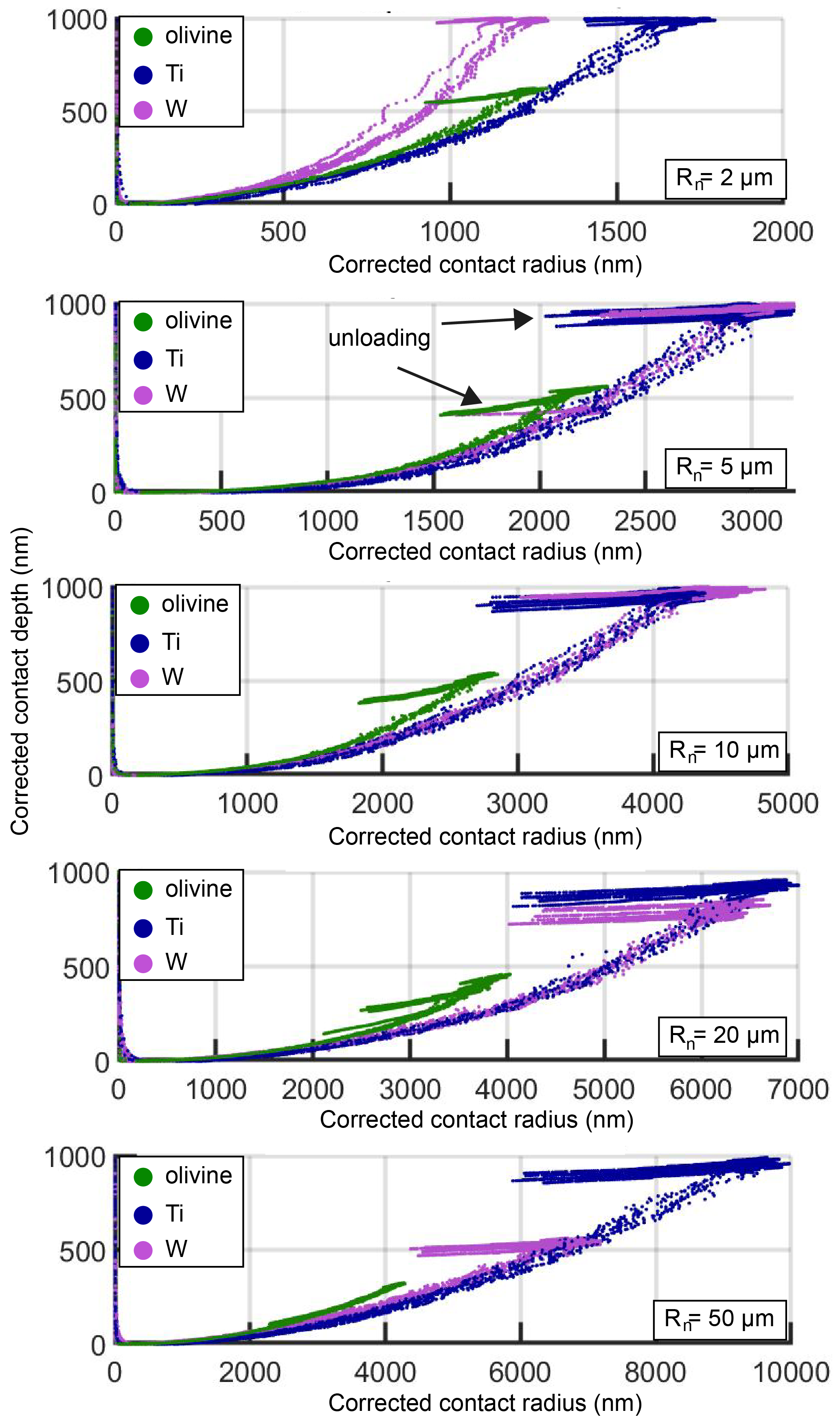}
\caption{Effective tip shape calculated with corrected data in olivine, tungsten, and titanium, with tips with varying radii. Note the differences in the horizontal axis. For largest tip radii, the indent experiments in olivine are purely elastic, whereas the residual imprint increases with decreasing radius, as displayed by the segment corresponding to the unloading. }
\label{fig:Supplementary1-methods}     
\end{figure*}

\begin{table*}[ht]
\centering
\caption{Summary of results presented in Figure \ref{figure6}. For each nominal radius $R_\mathrm{n}$ the Table presents $R_\mathrm{eff}$ calculated similar to \citet{li2013effects} using data collected in fused silica (FS), glassy carbon (GC), and sapphire (SPH) and $R_\mathrm{eff}$ calculated as a function of load as outlined in section \ref{radius_calculation}.}
\label{table_resultsSummary}       
\begin{tabular}{|c|ccc|cc|}
\hline\noalign{\smallskip}
{$R_\mathrm{n}$ \textnormal{\mu}m} & \multicolumn{3}{|c|}{$R_\mathrm{eff}$ \textnormal{\mu}m after \citet{li2013effects}} & \multicolumn{2}{|c|}{$R_\mathrm{eff}(P)$} \\ 
{}&{FS - SPH} & {GC - FS} & {GC-SPH} & $R_\mathrm{eff}$ \textnormal{\mu}m & P (N) \\
\noalign{\smallskip}\hline\noalign{\smallskip}
{2}&{0.3}&{0.25}&{0.41}&{0.76 $\pm$ 0.63}&{$2.5*10^{-3}$}\\
{5}&{6.97}&{6.97}&{8.4}&{3.43 $\pm$ 0.63}&{0.07}\\
{10}&{10.36}&{9.04}&{10.45}&{6.6 $\pm$ 0.8}&{0.08}\\
{20}&{23.5}&{32.3}&{28.75}&{18.2 $\pm$ 5.8}&{0.2}\\
{50}&{31.4}&{46}&{40}&{32 $\pm$ 4.7}&{0.25}\\
\noalign{\smallskip}\hline
\end{tabular}
\end{table*}

\begin{figure*}
  \includegraphics[width=0.9\textwidth]{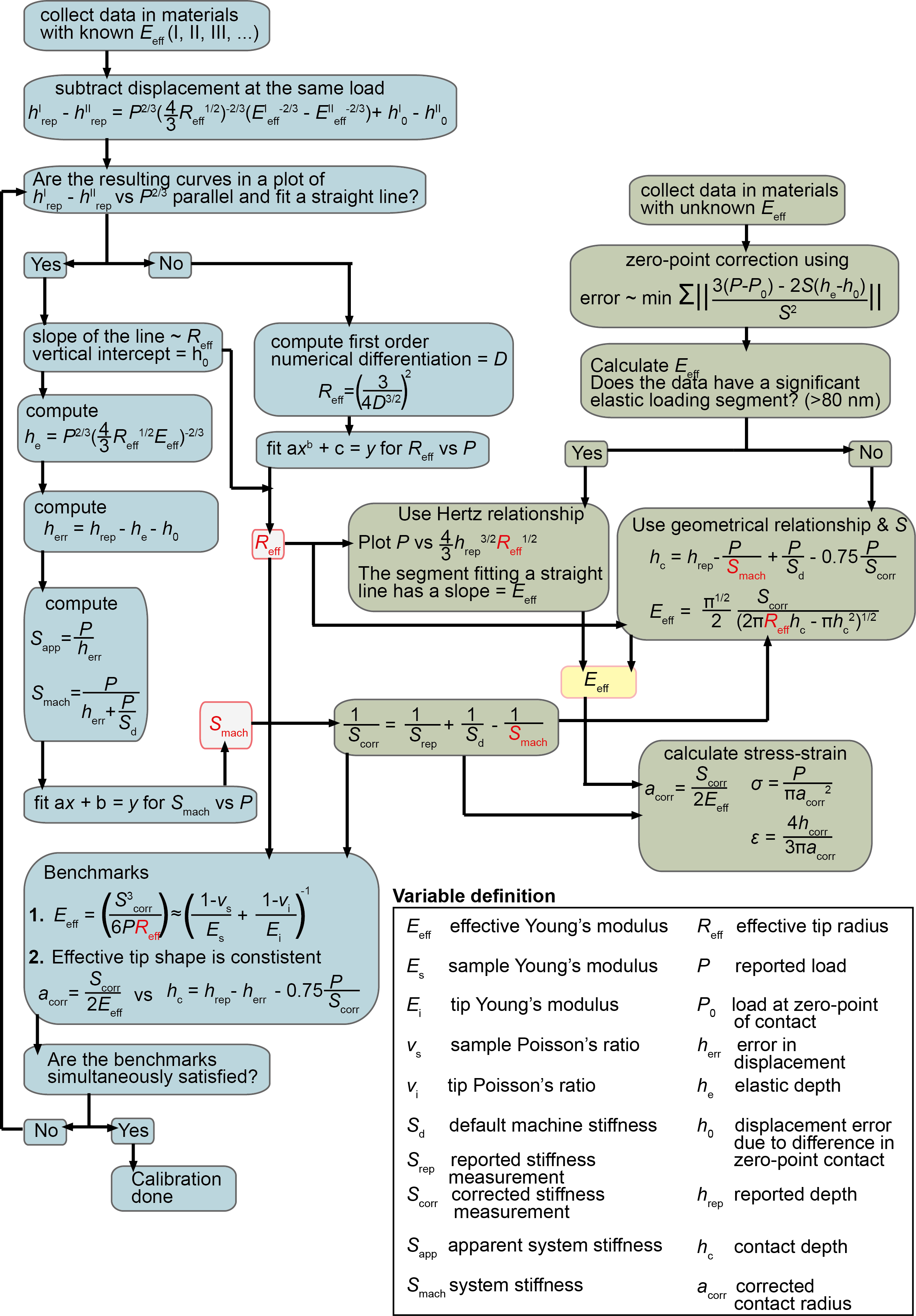}
\caption{Schematic flow-chart diagram summarising Section \ref{referenceMaterials} in blue and Section \ref{applicationMaterials} in green.}
\label{fig:Supplementary2-methods}     
\end{figure*}

\begin{figure*}
  \includegraphics[width=0.9\textwidth]{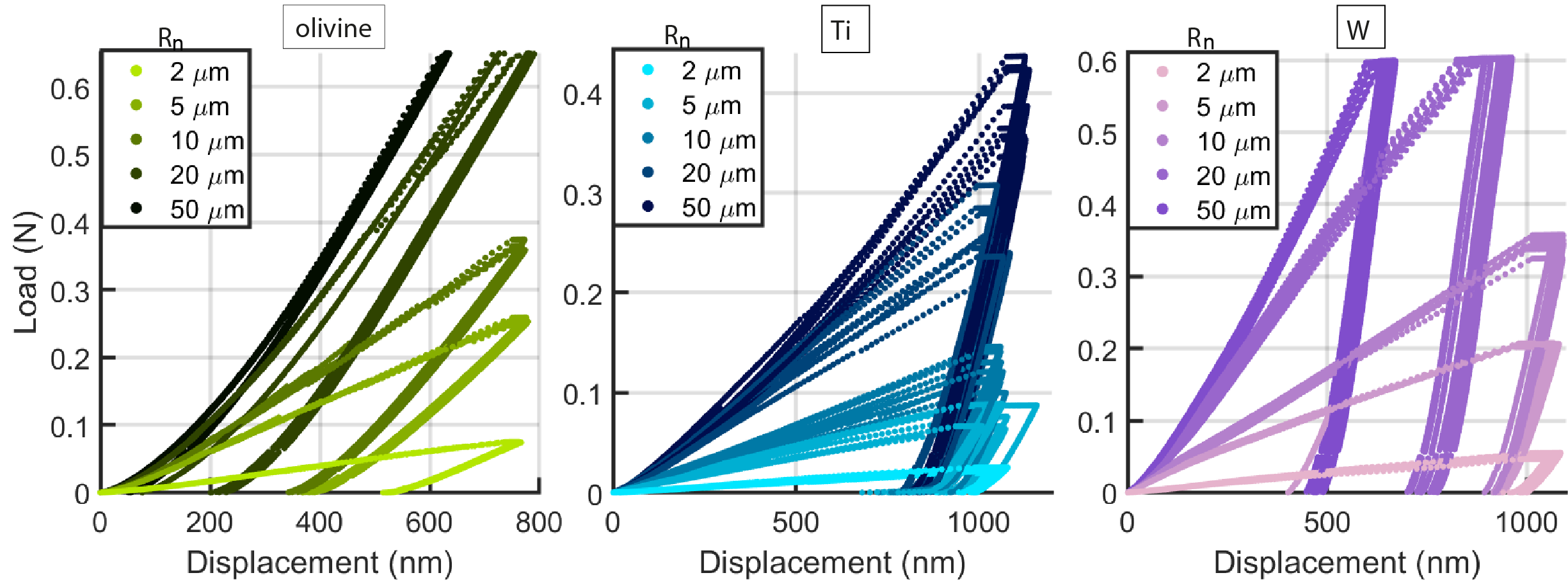}
\caption{Load-displacement curves in olivine, titanium, and tungsten collected during displacement-controlled experiments with spherical tips with nominal radii ranging $2-50$ \textnormal{\mu}m.}
\label{fig:Supplementary3-methods}     
\end{figure*}
%
%


\end{document}